\documentclass[aps,prb,reprint,twocolumn,notitlepage,superscriptaddress]{revtex4-1}
\usepackage{graphicx}
\usepackage{amsmath,amssymb}
\usepackage{color}
\usepackage{bm}
\usepackage{bbm}
\usepackage{mathtools}
\usepackage{hyperref}

\begin{document}

\title{Quasi-local Frustration-Free Free Fermions}
\author{Shunsuke Sengoku}
\affiliation{Department of Applied Physics, The University of Tokyo, Tokyo 113-8656, Japan}
\author{Hoi Chun Po}
\affiliation{Department of Physics, Hong Kong University of Science and Technology, Clear Water Bay, Hong Kong, China}
\affiliation{Center for Theoretical Condensed Matter Physics, \\
	Hong Kong University of Science and Technology, Clear Water Bay, Hong Kong, China}
\author{Haruki Watanabe}\email{hwatanabe@g.ecc.u-tokyo.ac.jp}
\affiliation{Department of Applied Physics, The University of Tokyo, Tokyo 113-8656, Japan}
\date{\today}

\begin{abstract}
Recent studies have revealed that frustration-free models, expressed as sums of finite-range interactions or hoppings, exhibit several properties markedly different from those of frustrated models. In this work, we demonstrate that, by relaxing the finite-range condition to allow for exponentially decaying hoppings, one can build  gapped frustration-free systems that realize Chern insulators as well as quasi-degenerate ground states with finite-size splittings. Moreover, by permitting power-law decaying hoppings, we  also construct a gapless band metal whose finite-size gap scales inversely with the system size $L$. 
These findings serve as an important step toward clarifying the general properties of frustration-free systems and those represented by tensor network states.
\end{abstract}
\maketitle

\section{Introduction}

Frustration-free models have played a pivotal role in theoretical studies of quantum many-body systems, owing to the relative ease with which their ground-state properties can be analyzed. Such models have deepened our understanding of diverse phenomena, including symmetry-broken phases~\cite{10.1063/1.1664978, Majumdar:1970aa,Mielke2,Mielke3,Tasaki}, symmetry-protected topological phases~\cite{AKLT, TasakiBook,PhysRevB.92.115137,PhysRevB.98.155119,Jevtic_2017}, and topologically ordered phases~\cite{KITAEV20032,PhysRevB.71.045110}. A system is said to be frustration-free if its Hamiltonian can be expressed as a sum
\begin{align}
  \hat{H} = \sum_i \hat{H}_i,
\end{align}
such that the ground state of the total Hamiltonian $\hat{H}$ simultaneously minimizes each individual term $\hat{H}_i$~\cite{TasakiBook}.

Recent investigations have uncovered several distinctive properties that arise when each $\hat{H}_i$ involves only finite-range interactions and hoppings. For instance, in gapped systems, ground-state degeneracies often remain exact even at finite system sizes~\cite{arXiv:2406.06414}. In gapless systems, the finite-size energy gap typically scales at least as quickly as $1/L^2$, where $L$ denotes the linear system size~\cite{PhysRevLett.120.117202,PhysRevB.103.214428,PhysRevResearch.3.033265,PhysRevLett.131.220403,arXiv:2405.00785,arXiv:2310.16881,arXiv:2406.06414}. This statement has been rigorously proven under open boundary conditions~\cite{Gossetozgunov,Anshu,Lemm_2022,lemm2024criticalfinitesizegapscaling}. However, it remains unclear whether these properties persist when the local terms $\hat{H}_i$ are allowed to include longer-range tails, such as exponentially or algebraically decaying interactions.

Another important feature of frustration-free systems concerns their topological properties. 
It is known that local commuting projector Hamiltonians acting on finite-dimensional Hilbert spaces cannot host chiral topological phases, such as those with thermal Hall conductance or (fractional) quantized Hall conductance~\cite{KITAEV20032,Kapustin,PhysRevB.105.L081103}. Although free-fermion chiral phases can be realized by Gaussian fermionic projected entangled-pair states (PEPS), 
these PEPS exhibit power-law correlations, and their parent Hamiltonians are necessarily either gapless or nonlocal with power-law hopping tails~\cite{PhysRevLett.111.236805,PhysRevB.90.115133},\footnote{For the example in Ref.~\onlinecite{PhysRevB.90.115133}, the power-law exponent is $3$.}
reflecting the fact that Gaussian fermionic tensor network states with local, gapped parent Hamiltonians cannot realize chiral phases~\cite{PhysRevB.92.205307}. This is closely related to the absence of compactly supported Wannier functions in Chern insulators~\cite{RevModPhys.84.1419,PhysRevB.95.115309}.

These observations have led to the following conjecture: \emph{no gapped frustration-free Hamiltonian can realize a chiral phase}. 
Evidence for this has been found in free-fermion frustration-free systems with strictly short-range interactions~\cite{Ono2025FrustrationFree,Masaoka2025FrustrationFree}, and further supported by the incompatibility between chiral edge modes with linear dispersion (implying a finite-size gap of order $L^{-1}$) and the general $O(L^{-2})$ scaling of the finite-size gap in gapless frustration-free systems~\cite{10.1063/1.5089773}.
However, it remains open 
whether a gapped frustration-free Hamiltonian with exponentially decaying hoppings and/or many-body interactions under periodic boundary conditions could host a chiral phase.

In this work, we address how the relaxation of the assumption of strictly finite-range terms in the frustration-free decomposition affects the scaling property of the excitation gap and the representability of gapped chiral phases.
We show that allowing each local term $\hat{H}_i$ to have exponentially decaying tails enables the realization of gapped, frustration-free free-fermion models exhibiting nontrivial band topology. These include models with nonzero Chern numbers and edge states, leading to quasi-degenerate ground states with finite-size splittings. A similar construction was demonstrated recently in Ref.~\onlinecite{PhysRevResearch.3.033265} for the transverse-field Ising model, showing finite-size splitting in a sequence of exponentially decaying frustration-free Hamiltonians. Moreover, extending the decay profile to algebraic (power-law) forms allows us to construct gapless models whose finite-size gaps scale as $1/L$. Crucially, while the individual terms $\hat{H}_i$ may exhibit exponential or power-law decay, the total Hamiltonian $\hat{H}=\sum_i \hat{H}_i$ itself remains strictly local, consisting solely of finite-range hoppings.

These findings underscore the sharp distinction between frustration-free decompositions with strictly finite-range terms and those permitting longer-range tails. Our results thereby sharpen the aforementioned conjecture into 
\emph{no gapped frustration-free Hamiltonian $\hat{H}=\sum_i \hat{H}_i$ can realize a chiral phase if each local term $\hat{H}_i$ in the frustration-free decomposition is finite-ranged}. As such, we provide a framework for investigating more general classes of frustration-free systems—including those described by tensor network states such as PEPS—and highlight the role of extended hopping tails in circumventing previously assumed constraints.

\section{Formulation}
\subsection{Tight-binding model}
In this section, we consider a general tight-binding model of the following form:
\begin{align}
\hat{H}\coloneqq\sum_{i,j}\hat{c}_i^\dagger h_{i,j} \hat{c}_j=\hat{\bm{c}}^\dagger h \hat{\bm{c}},\label{h1}
\end{align}
where $h$ is the matrix whose elements are the hopping amplitudes $h_{i,j}$. This matrix can be diagonalized by a unitary matrix $U$ as
\begin{align}
h_{i,j}=\sum_{n}U_{i,n}\epsilon_n U_{j,n}^*.
\end{align}
Using the projector 
\begin{align}
(P_n)_{i,j}\coloneqq U_{i,n}U_{j,n}^* \label{pj}
\end{align}
onto the eigenspace corresponding to $\epsilon_n$, we can decompose $h$ into its positive-energy and negative-energy components:
\begin{align}
&h=\sum_{n}\epsilon_nP_n=h^{(+)}+h^{(-)},\\
&h^{(\pm)}\coloneqq \sum_{n|\pm\epsilon_n>0}\epsilon_nP_n,\label{hpm}
\end{align}
so that the Hamiltonian in Eq.~\eqref{h1} can also be expressed as
\begin{align}
\hat{H}&=\hat{\bm{c}}^\dagger h^{(+)}\hat{\bm{c}}+\hat{\bm{c}}^\dagger h^{(-)}\hat{\bm{c}}\notag\\
&=\hat{\bm{c}}^\dagger h^{(+)}\hat{\bm{c}}+\hat{\bm{c}}^T(-h^{(-)}{}^T)\bigl(\hat{\bm{c}}^\dagger\bigr)^T+E_0,\label{h2}
\end{align}
where $E_0\coloneqq\mathrm{tr}[h^{(-)}]=\sum_{\epsilon_n<0}\epsilon_n$ is the ground-state energy.

To write down the ground state $|\Phi\rangle$, we introduce the creation operators for eigenmodes
\begin{align}
\hat{\gamma}_n^\dagger\coloneqq\sum_i\hat{c}_i^\dagger U_{i,n}.
\end{align}
The ground state is the fully occupied state of the negative-energy levels:
\begin{align}
|\Phi\rangle\coloneqq\prod_{\epsilon_n<0}\hat{\gamma}_n^\dagger|0\rangle.
\end{align}
Here and hereafter, we assume that $h$ has no eigenvalue exactly equal to zero.\footnote{If such an eigenvalue exists, the degeneracy of the ground state must be taken into account.} 
The projector in Eq.~\eqref{pj} is related to the ground-state two-point functions:
\begin{align}
\langle\Phi|\hat{c}_i\hat{c}_j^\dagger|\Phi\rangle
&=\sum_{n}U_{j,n}^*\langle\Phi|\hat{\gamma}_n\hat{\gamma}_n^\dagger|\Phi\rangle U_{i,n}
=\sum_{\epsilon_n>0}(P_n)_{i,j},\\
\langle\Phi|\hat{c}_j^\dagger\hat{c}_i|\Phi\rangle
&=\sum_{n}U_{j,n}^*\langle\Phi|\hat{\gamma}_n^\dagger\hat{\gamma}_n|\Phi\rangle U_{i,n}
=\sum_{\epsilon_n<0}(P_n)_{i,j}.
\end{align}

\subsection{Frustration-free decomposition}
Next, we convert this Hamiltonian to a frustration-free form. In Appendix D.1.2 of Ref.~\onlinecite{KITAEV20062}, the case of quadratic Hamiltonians of Majorana fermions is discussed. Here, we translate that argument into tight-biding models of complex fermions.

To this end, note that $+h^{(+)}$ and $-h^{(-)}$ in Eq.~\eqref{hpm} are both positive-definite. Hence, their square-root 
\begin{align}
\sqrt{\pm h^{(\pm)}}\coloneqq\sum_{n|\pm\epsilon_n>0}\sqrt{|\epsilon_n|}P_n \label{sr}
\end{align}
is well-defined. We can rewrite the Hamiltonian in Eq.~\eqref{h2} as
\begin{align}
\hat{H}&=\hat{\bm{c}}^\dagger \sqrt{h^{(+)}}\sqrt{h^{(+)}}\hat{\bm{c}}\notag\\
&\quad+\hat{\bm{c}}^T\sqrt{-h^{(-)}{}^T}\sqrt{-h^{(-)}{}^T}(\hat{\bm{c}}^\dagger)^T +E_0\notag\\
&=\sum_i\hat{H}_i+E_0,\label{h3}\\
\hat{H}_i&\coloneqq(\hat{\psi}_i^{(+)})^\dagger\hat{\psi}_i^{(+)}+\hat{\psi}_i^{(-)}(\hat{\psi}_i^{(-)})^\dagger,
\end{align}
where 
\begin{align}
&\hat{\psi}_i^{(\pm)}\coloneqq \sum_j\Big(\sqrt{\pm h^{(\pm)}}\Big)_{i,j}\hat{c}_j=\sum_{n|\pm\epsilon_n>0}\sqrt{|\epsilon_n|}U_{i,n}\hat{\gamma}_n.
\end{align}
Clearly, $\hat{H}_i$ is a sum of positive-semidefinite terms. Moreover, both $\hat{\psi}_i^{(+)}$ and $\big(\hat{\psi}_i^{(-)}\big)^\dagger$ annihilate the ground state $|\Phi\rangle$ because $\hat{\psi}_i^{(+)}$ is a sum of $\hat{\gamma}_n$ of positive energy states and $\big(\hat{\psi}_i^{(-)}\big)^\dagger$ is a sum of $\hat{\gamma}_n^\dagger$ of negative energy states. Therefore, 
\begin{align}
\hat{H}_i|\Phi\rangle=0
\end{align}
for all $i$ simultaneously, implying that the Hamiltonian with the decomposition in Eq.~\eqref{h3} is frustration-free.

The square-root of the matrices in Eq.~\eqref{sr} can also be represented by the ground-state correlation functions as
\begin{align}
\Big(\sqrt{+ h^{(+)}}\Big)_{i,j}=\langle\Phi|\hat{c}_i\sqrt{\hat{H}-E_0}\,\hat{c}_j^\dagger|\Phi\rangle,\\
\Big(\sqrt{- h^{(-)}}\Big)_{i,j}=\langle\Phi|\hat{c}_j^\dagger\sqrt{\hat{H}-E_0}\,\hat{c}_i|\Phi\rangle.
\end{align}

\subsection{Exponential localization}
In general, $\hat{H}_i$ constructed above is not necessarily local. However, if $\hat{H}$ is local and its spectrum is gapped, each $\hat{H}_i$ becomes quasi-local with exponentially decaying tail.

To state this result more precisely, suppose that $\hat{H}$ is local in the sense that 
\begin{align}
|h_{i,j}|<C_0 e^{-A_0 d(i-j)}
\end{align}
for positive constants $C_0$ and $A_0$, where $d(i,j)$ denotes the distance between the indices $i$ and $j$. 
In one-dimensional models with open boundary conditions, one may simply set $d(i,j)=|i-j|$, but we are interested in more general settings. 
Furthermore, assume there is a gap
\begin{align}
\Delta\coloneqq\min_{n|\epsilon_n>0}\{\epsilon_n\}-\max_{n|\epsilon_n<0}\{\epsilon_n\}>0
\end{align}
between the positive and negative eigenvalues, and that this gap remains nonzero in the limit of large system size. Then one can show
\begin{align}\label{eq:exponential_square_root}
\big|\big(\sqrt{\pm h^{(\pm)}}\big)_{i,j}\big|<C e^{-A d(i-j)}
\end{align}
for some constants $C$ and $A$, implying that each $\hat{\psi}_i^{(\pm)}$ is exponentially localized. 
As we discuss below in more details for the example of a Chern insulator, when the system is translation-invariant the asserted exponential decay in Eq.\ \eqref{eq:exponential_square_root} follows from the analyticity of its momentum-space counterpart when it is extended to complex momenta.
We will see this exponential decay in more detail for band insulators.

\section{Gapless model with $O(1/L)$ finite-size gap}
In this section, we discuss a gapless frustration-free system that exhibits an $O(1/L)$ finite-size gap.

\begin{figure}[t]
\begin{center}
\includegraphics[width=\columnwidth]{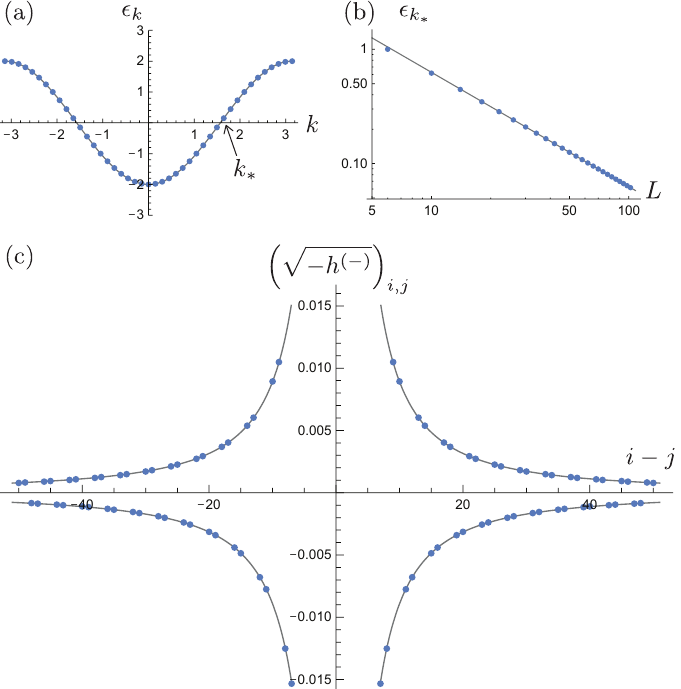}
\end{center}
\caption{
Results for the one-band model in Eq.~\eqref{ex1} with $t=1$.
(a) The band dispersion $\epsilon_k=-2\cos k$ for $L=42$.
(b) The finite-size gap $\epsilon_{k_*}=2\sin(\pi/L)$ with $k_*\coloneqq\frac{\pi}{2}+\frac{\pi}{L}$. The solid line indicates $2\pi/L$.
(c) The plot of $\big(\sqrt{- h^{(-)}}\big)_{i,j}$ as a function of $i-j$ for $L=1002$. The solid curve represents $(4\pi)^{-1/2}|i-j|^{-3/2}$ for comparison.
\label{fig1}
}
\end{figure}

\subsection{One-band model}
We consider the simplest one-dimensional tight-binding model:
\begin{align}
\hat{H}=-t\sum_{i=1}^L\big(\hat{c}_{i+1}^\dagger\hat{c}_{i}+\hat{c}_{i}^\dagger\hat{c}_{i+1}\big)\quad (t>0),\label{ex1}
\end{align}
which can be solved by the Fourier transformation $\hat{c}_k\coloneqq\frac{1}{\sqrt{L}}\sum_{j=1}^Le^{-ikj}\hat{c}_j$.
The band dispersion $\epsilon_k=-2t\cos k$ is negative for $|k|<\frac{\pi}{2}$ and positive for $\frac{\pi}{2}<|k|<\pi$ as illustrated in Fig.~\ref{fig1}(a).  To avoid a zero-energy eigenstate, we choose $L$ \emph{not} to be an integer multiple of four. 

The ground state is the fully occupied state of negative energy states and the vacuum state of positive energy states:
\begin{align}
|\Phi\rangle=\prod_{-\frac{\pi}{2}<k<\frac{\pi}{2}}\hat{c}_k^\dagger|0\rangle.
\end{align}
The ground state energy is $E_0=\sum_{-\frac{\pi}{2}<k<\frac{\pi}{2}}\epsilon_k$.

In this model, excitations are gapless, and the finite-size gap scales as $O(L^{-1})$. For example, when the system size is $L=4m+2$, the smallest excitation energy occurs at at $k_*\coloneqq\frac{\pi}{2}+\frac{\pi}{L}$ [see Fig.~\ref{fig1}(a,b)]
\begin{align}
\epsilon_{k_*}=2t\sin\Big(\frac{\pi}{L}\Big)=O(L^{-1}).
\end{align}
\subsection{Frustration-free decomposition}
The Hamiltonian \eqref{ex1} can be written in a frustration-free form:
\begin{align}
&\hat{H}=\sum_{i}\hat{H}_i+E_0=\sum_{k}\epsilon_k\,\hat{c}_k^\dagger\hat{c}_{k},\\
&\hat{H}_j\coloneqq(\hat{\psi}_j^{(+)})^\dagger\hat{\psi}_j^{(+)}+\hat{\psi}_j^{(-)}(\hat{\psi}_j^{(-)})^\dagger,
\end{align}
where
\begin{align}
\hat{\psi}_j^{(\pm)}&\coloneqq\sum_{j'}\Big(\sqrt{\pm h^{(\pm)}}\Big)_{j,j'}\hat{c}_{j'}=\frac{1}{\sqrt{L}}\sum_{k|\pm\epsilon_k>0}\sqrt{|\epsilon_k|}\,e^{ikj}\hat{c}_k.
\end{align}
Because $\hat{\psi}_j^{(+)}$ and $(\hat{\psi}_j^{(-)})^\dagger$ annihilate the ground state $|\Phi\rangle$, it follows that $\hat{H}_j|\Phi\rangle=0$ for all $j$ and the Hamiltonian is frustration-free under this decomposition.

In this model, excitations are not gapped and
\begin{align}
\Big(\sqrt{\pm h^{(\pm)}}\Big)_{i,j}&\coloneqq\frac{1}{L}\sum_{k|\pm\epsilon_k>0}\sqrt{|\epsilon_k|}\,e^{ik(i-j)}
\end{align}
do not decay exponentially in $|i-j|$.  Instead, we find a power-law decay of order $|i-j|^{-3/2}$:
\begin{align}
\Big(\sqrt{h^{(+)}}\Big)_{i,j}&=\Big(\sqrt{-h^{(-)}}\Big)_{i,j}\,e^{i\pi (i-j)},\\
\Big(\sqrt{-h^{(-)}}\Big)_{i,j}&\simeq\int_{-\pi/2}^{\pi/2}\frac{dk}{2\pi}\sqrt{2t\cos k}\,e^{ik(i-j)}\notag\\
&=\frac{\sqrt{t\pi}}{4}\frac{1}{\Gamma(\frac{5}{4}-\frac{i-j}{2})\Gamma(\frac{5}{4}+\frac{i-j}{2})}\notag\\
&\simeq\sqrt{\frac{t}{4\pi}}\frac{(-1)^{\lfloor\frac{|i-j|-1}{2}\rfloor}}{|i-j|^{3/2}}.
\end{align}
Here, $\Gamma(z)$ is the gamma function and $\lfloor x\rfloor$ denotes the greatest integer not exceeding $x$. In deriving the final expression, we replaced the sum over $k$ by an integral, which becomes exact in the large-$L$ limit. We then used the reflection formula $\Gamma(z)\Gamma\bigl(1-z\bigr)=\pi/\sin(\pi z)$ and Stirling's approximation $\Gamma(z)\simeq\sqrt{\tfrac{2\pi}{z}}\bigl(\tfrac{z}{e}\bigr)^z$. A numerical check of this power-law decay is shown in Fig.~\ref{fig1}(c).

\section{Chern insulator}
Next, we consider an example of a frustration-free band insulator with a nonzero Chern number.

\subsection{Qi--Wu--Zhang model}
We consider the Qi--Wu--Zhang model~\cite{PhysRevB.74.085308} on a square lattice, where each lattice site is located at $\bm{R} = (x, y)$ with $x, y = 1, 2, \ldots, L$.
\begin{align}
\hat{H}=
&\sum_{\bm{R}}\begin{pmatrix}\hat{c}_{\bm{R}+\bm{e}_x1}^\dagger&\hat{c}_{\bm{R}+\bm{e}_x2}^\dagger\end{pmatrix}t\Big(\frac{i}{2} \sigma_1-\frac{1}{2}\sigma_3\Big)\begin{pmatrix}\hat{c}_{\bm{R}1}\\\hat{c}_{\bm{R}2}\end{pmatrix}+\text{h.c.}\notag\\
&\quad+\sum_{\bm{R}}\begin{pmatrix}\hat{c}_{\bm{R}+\bm{e}_y1}^\dagger&\hat{c}_{\bm{R}+\bm{e}_y2}^\dagger\end{pmatrix}t\Big(\frac{i}{2} \sigma_2-\frac{1}{2}\sigma_3\Big)\begin{pmatrix}\hat{c}_{\bm{R}1}\\\hat{c}_{\bm{R}2}\end{pmatrix}+\text{h.c.}\notag\\
&\quad+\sum_{\bm{R}}\begin{pmatrix}\hat{c}_{\bm{R}1}^\dagger&\hat{c}_{\bm{R}2}^\dagger\end{pmatrix}
m\sigma_3\begin{pmatrix}\hat{c}_{\bm{R}1}\\\hat{c}_{\bm{R}2}\end{pmatrix}\quad(t>0).\label{Chern}
\end{align}
Here, $\sigma_1$, $\sigma_2$, and $\sigma_3$ denote the Pauli matrices, and $\bm{e}_x$ and $\bm{e}_y$ are the unit vectors along the $x$- and $y$-axes, respectively. For simplicity, we set $t = 1$ in the following.

\begin{figure}[t]
\begin{center}
\includegraphics[width=\columnwidth]{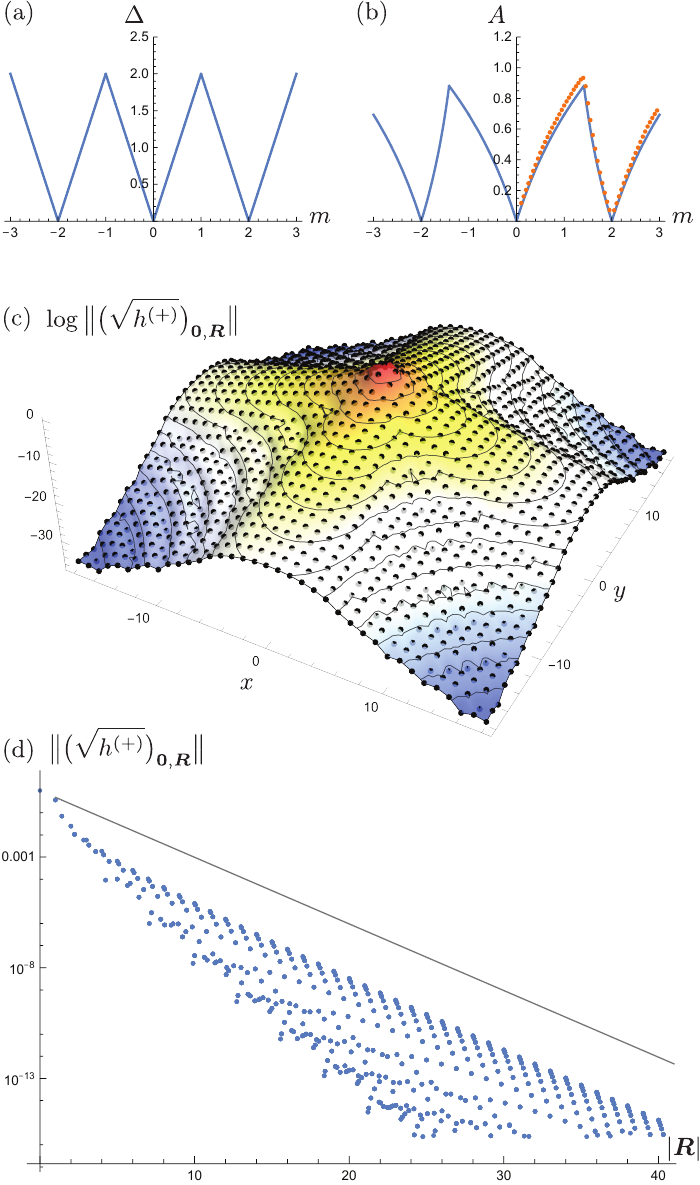}
\end{center}
\caption{
Results on the Qi-Wu--Zhang model in Eq.~\eqref{Chern} with $t=1$.
(a) The band gap $\Delta$ as a function of $m$.
(b) $A$ in Eq.~\eqref{Achern} as a function of $m$. Orange dots represent values obtained by fitting data for $L=201$ and $0<m<3$. 
(c,d) The plot of $\big\|\big(\sqrt{h^{(+)}}\big)_{\bm{0},\bm{R}}\big\|\coloneqq\sqrt{\sum_{\sigma,\sigma'}\big|\big(\sqrt{h^{(+)}}\big)_{\bm{0}\sigma,\bm{R}\sigma'}\big|^2}$ as a function of $\bm{R}=(x,y)$ for $m=1$ and $L=101$.
The line in (d) represents $e^{-A|\bm{R}|}$ with $A=\log2$.
\label{fig2}
}
\end{figure}

After the Fourier transformation $\hat{c}_{\bm{k}\sigma}^\dagger\coloneqq\frac{1}{L}\sum_{\bm{R}}\hat{c}_{\bm{R}\sigma}^\dagger e^{i\bm{k}\cdot\bm{R}}$, the Hamiltonian becomes
\begin{align}
\hat{H}=
\begin{pmatrix}\hat{c}_{\bm{k}1}^\dagger&\hat{c}_{\bm{k}2}^\dagger\end{pmatrix}
h_{\bm{k}}
\begin{pmatrix}\hat{c}_{\bm{k}1}\\\hat{c}_{\bm{k}2}\end{pmatrix}=\sum_{\sigma,\sigma'=1}^2\hat{c}_{\bm{k}\sigma}^\dagger(h_{\bm{k}})_{\sigma,\sigma'}\hat{c}_{\bm{k}\sigma'},
\end{align}
where $h_{\bm{k}}$ can be expanded by the Pauli matrices as
\begin{align}
h_{\bm{k}}=\sum_{\alpha=1}^3g_{\alpha\bm{k}}\sigma_\alpha\label{hmatrix}
\end{align}
 with $g_{1\bm{k}}=\sin k_x$, $g_{2\bm{k}}=\sin k_y$, and $g_{3\bm{k}}=m-\cos k_x-\cos k_y$. The eigenvalues of $h_{\bm{k}}$ are $\pm\epsilon_{\bm{k}}$ with $\epsilon_{\bm{k}}\coloneqq\sqrt{\sum_{\alpha=1}^3g_{\alpha\bm{k}}^2}$.
The ground state $|\Phi\rangle$ is the fully occupied state of the lower band $-\epsilon_{\bm{k}}$, and the ground-state energy is 
$E_0=-\sum_{\bm{k}}\epsilon_{\bm{k}}$.
Excitations are gapped unless $m=0,2,-2$, as shown in Fig.~\ref{fig2}(a).

The Chern number of the lower-band 
\begin{align}
C=\int \frac{dk_xdk_y}{2\pi}\frac{\cos k_x+\cos k_y-m\cos k_x\cos k_y}{2 \epsilon_k^3}
\end{align}
is given by $C=\mathrm{sign}(m)$ for $0<|m|<2$, and $C=0$ for $2<|m|$. 

\subsection{Frustration-free decomposition}
Let us decompose the Hamiltonian into a frustration-free form.
To this end, we note that the projector onto the eigenvalue $\pm\epsilon_{\bm{k}}$ of $h_{\bm{k}}$ can be written as
\begin{align}
P_{\bm{k},\pm}=\frac{1}{2}\sigma_0\pm\frac{h_{\bm{k}}}{2\epsilon_{\bm{k}}}.
\end{align}
Using this expression, we define
\begin{align}
&(h^{(\pm)})_{\bm{R}\sigma,\bm{R}'\sigma'}\coloneqq\frac{1}{L^2}\sum_{\bm{k}}(\pm\epsilon_{\bm{k}})(P_{\bm{k},\pm})_{\sigma,\sigma'}e^{i\bm{k}\cdot(\bm{R}-\bm{R}')}.\label{sum}
\end{align}
Then the Hamiltonian in Eq.~\eqref{Chern} can be rewritten as
\begin{align}
&\hat{H}=\sum_{\bm{R}}\hat{H}_{\bm{R}}+E_0,\\
&\hat{H}_{\bm{R}}\coloneqq\sum_{\sigma=1}^2\Big(\big(\hat{\psi}_{\bm{R}\sigma}^{(+)}\big)^\dagger\hat{\psi}_{\bm{R}\sigma}^{(+)}+
\hat{\psi}_{\bm{R}\sigma}^{(-)}\big(\hat{\psi}_{\bm{R}\sigma}^{(-)}\big)^\dagger
\Big),\\
&\hat{\psi}_{\bm{R}\sigma}^{(\pm)}\coloneqq \sum_{\bm{R}'\sigma'}\Big(\sqrt{\pm h^{(\pm)}}\Big)_{\bm{R}\sigma,\bm{R}'\sigma'}\hat{c}_{\bm{R}'\sigma'}.
\end{align}
Clearly, $\hat{H}_{\bm{R}}$ is the sum of positive-semidefinite terms. Also, $\hat{H}_{\bm{R}}|\Phi\rangle=0$ since $\hat{\psi}_{\bm{R}\sigma}^{(+)}|\Phi\rangle=\big(\hat{\psi}_{\bm{R}\sigma}^{(-)}\big)^\dagger|\Phi\rangle=0$. Therefore, the Hamiltonian with this decomposition is frustration-free.

Now we show that $\hat{H}_{\bm{R}}$ has an exponentially decaying tail.
Since $\epsilon_{\bm{k}}>0$ for any $\bm{k}$, the quantity $\sqrt{\epsilon_{\bm{k}}}P_{\bm{k},\pm}$ is an analytic function of $\bm{k}$.
Moreover, 
we can extend $\bm{k}$ to complex values, and $\sqrt{\epsilon_{\bm{k}}}P_{\bm{k},\pm}$ remains analytic in $\bm{k}\in\mathbb{C}^d$ within the domain $\sum_{\alpha=1}^d[\mathrm{Im}(k_\alpha)]^2<A^2$. Here $A>0$ defined by
\begin{align}
&A\coloneqq
\begin{cases}
\log(|m|+1)&(\sqrt{2}>|m|>0)\\
|\log(|m|-1)|&(|m|>\sqrt{2})
\end{cases}.\label{Achern}
\end{align}
and Fig.~\ref{fig2}(b) shows that $A$ depends on $m$ in a way similar to the excitation gap in Fig.~\ref{fig2}(a).
It follows that the matrix elements exhibit exponential decay of the form
\begin{align}
\Big|\Big(\sqrt{\pm h^{(\pm)}}\Big)_{\bm{R}\sigma,\bm{R}'\sigma'}\Big|<Ce^{-A|\bm{R}-\bm{R}'|}\label{expdecayChern}
\end{align}
for a constant $C>0$~\footnote{Mathematically this result is known as the Paley--Wiener theorem. See Sec~4 and Sec~7.1 of Ref.~\onlinecite{katznelson2004harmonic} for details.
}. This can be seen by
approximating the sum in Eq.~\eqref{sum} by an integral
\begin{align}
\Big(\sqrt{\pm h^{(\pm)}}\Big)_{\bm{R}\sigma,\bm{R}'\sigma'}\simeq\int\frac{d^dk}{(2\pi)^d}\sqrt{\epsilon_{\bm{k}}}(P_{\bm{k},\pm})_{\sigma,\sigma'}e^{i\bm{k}\cdot(\bm{R}-\bm{R}')}
\end{align}
and then shifting $\bm{k}$ to $\bm{k}+iA\frac{\bm{R}}{|\bm{R}|}$ using Cauchy's integral theorem. 
This establishes the exponential locality of $\hat{\psi}_{\bm{R}\sigma}^{(\pm)}$.
We numerically verify these results in Fig.~\ref{fig2}(c,d).

In contrast to the chiral PEPS and their parent Hamiltonians discussed in Refs.~\onlinecite{PhysRevLett.111.236805,PhysRevB.90.115133},
the ground-state correlations and hoppings of our Hamiltonian decay exponentially rather than following a power law.

\section{Finite-size splitting of quasi-degenerate ground states}
Finally, we consider a gapped frustration-free Hamiltonian that exhibits degenerate ground states with finite-size splitting.

\subsection{Su--Schrieffer--Heeger model}
We consider the Su--Schrieffer--Heeger model~\cite{PhysRevLett.42.1698} under open boundary conditions:
\begin{align}
\hat{H}&=-\sum_{j=1}^{L}t\big(\hat{c}_{j+1,1}^\dagger\hat{c}_{j,2}+\hat{c}_{j,2}^\dagger\hat{c}_{j+1,1}\big)\notag\\
&\quad-\sum_{j=1}^{L}\mu\big(\hat{c}_{j,1}^\dagger\hat{c}_{j,1}+\hat{c}_{j,1}^\dagger\hat{c}_{j,2}\big)\label{SSH}
\end{align}
The model possesses a sublattice symmetry $\hat{\Gamma}$ defined by
\begin{align}
\hat{\Gamma}\hat{c}_{j,\sigma}\hat{\Gamma}^\dagger=(-1)^\sigma \hat{c}_{j,\sigma}.
\end{align}
We have 
\begin{align}
\hat{\Gamma}\hat{H}\hat{\Gamma}^\dagger=-\hat{H}.
\end{align}
When $0 < \mu\big(1+\frac{1}{L}\big) < t$, the system is topologically nontrivial and hosts an edge mode on each boundary. For finite $L$, these two edge modes hybridize, producing an exponentially small energy splitting.

\begin{figure}[t]
\begin{center}
\includegraphics[width=\columnwidth]{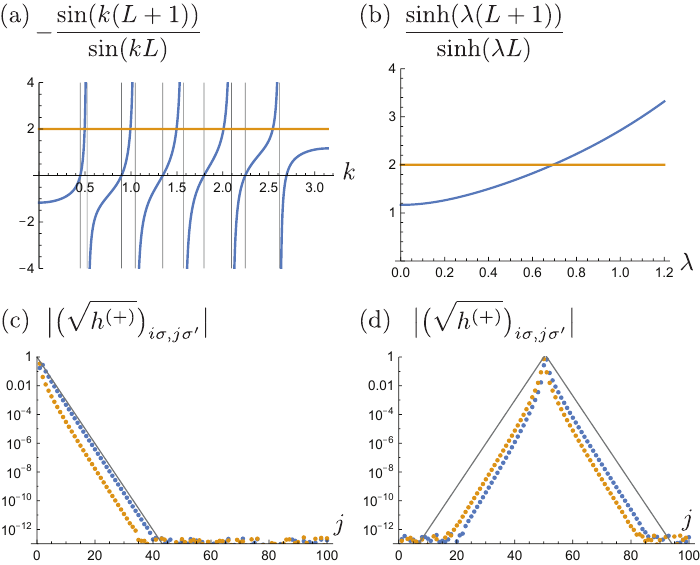}
\end{center}
\caption{
Results for the Su--Schrieffer--Heeger model in Eq.~\eqref{SSH} with $t=2$ and $\mu=1$
(a,b) Illustration of the solutions of Eq.~\eqref{eqkn} and Eq.~\eqref{eqlambda} for $L=6$.
(c,d) The plot of $\big\|\big(\sqrt{h^{(+)}}\big)_{\bm{0},\bm{R}}\big\|\coloneqq\sqrt{\sum_{\sigma,\sigma'}\big|\big(\sqrt{h^{(+)}}\big)_{i\sigma,j\sigma'}\big|^2}$ [$(i,\sigma)=(1,1)$ for (c) and $(i,\sigma)=(50,2)$ for $(d)$] as a function of $j$ for $L=101$. The blue (orange) dots correspond to $\sigma'=1$ ($\sigma'=2$). The slopes of the solid lines are $\pm\log(t/\mu)$.
\label{fig3}
}
\end{figure}

Following Ref.~\onlinecite{PhysRevB.95.195140}, we diagonalize the Hamiltonian exactly. When $\mu\big(1+\frac{1}{L}\big)<t$, we find
\begin{align}
\hat{H}&=\sum_{n=1}^{L-1}\epsilon_{n}(\hat{\gamma}_n^\dagger\hat{\gamma}_n-\hat{\tilde{\gamma}}_n^\dagger\hat{\tilde{\gamma}}_n)\notag\\
&\quad+\epsilon_{\text{edge}}(\hat{\gamma}_{\text{edge}}^\dagger\hat{\gamma}_{\text{edge}}-\hat{\tilde{\gamma}}_{\text{edge}}^\dagger\hat{\tilde{\gamma}}_{\text{edge}}),
\end{align}
where the $2(L-1)$-bulk modes are given by
\begin{align}
&\hat{\gamma}_{n}^\dagger\coloneqq\sum_{j=1}^L\sum_{\sigma=1}^2\phi_{j\sigma,n}\hat{c}_{j\sigma}^\dagger,\\
&\hat{\tilde{\gamma}}_{n}^\dagger\coloneqq\hat{\Gamma}\hat{\gamma}_{n}^\dagger\hat{\Gamma}^\dagger=\sum_{j=1}^L\sum_{\sigma=1}^2\phi_{j\sigma,n}(-1)^\sigma\hat{c}_{j\sigma}^\dagger,\\
&\phi_{j1,n}\coloneqq- N_n^{-1/2}(-1)^{n-1}\sin(k_n j),\\
&\phi_{j2,n}\coloneqq N_n^{-1/2}\sin(k_n (L-j+1)),\\
&N_n\coloneqq L+\frac{1}{2}-\frac{\sin k(2L+1)}{2\sin k},\\
&\epsilon_{n}\coloneqq\Big|\frac{\sin k_n}{\sin(k_nL)}\Big|=\sqrt{t^2+\mu^2+2t\mu\cos k_n}
\end{align}
and $k_n$ is the solution of
\begin{align}
-\frac{\sin(k(L+1))}{\sin(kL)}=\frac{t}{\mu}\label{eqkn}
\end{align}
in the range $\frac{\pi}{L+1}n<k<\frac{\pi}{L}n$ for $n=1,2,\cdots,L-1$ [Fig.~\ref{fig3}(a)]~\footnote{When $0<t<\mu(1+\frac{1}{L})$, there is an additional solution if $k$ in the range $\frac{\pi L}{L+1}<k<\pi$ instead of the edge mode.}. Similarly, the two edge modes are given by
\begin{align}
&\hat{\gamma}_{\text{edge}}^\dagger\coloneqq\sum_{j=1}^L\sum_{\sigma=1}^2\phi_{j\sigma,\text{edge}}\hat{c}_{j\sigma}^\dagger,\\
&\hat{\tilde{\gamma}}_{\text{edge}}^\dagger\coloneqq\hat{\Gamma}\hat{\gamma}_{n}^\dagger\hat{\Gamma}^\dagger=\sum_{j=1}^L\sum_{\sigma=1}^2\phi_{j\sigma,\text{edge}}(-1)^\sigma\hat{c}_{j\sigma}^\dagger,\\
&\phi_{j1,\text{edge}}\coloneqq - N_{\text{edge}}^{-1/2}(-1)^j\sinh(\lambda j),\\
&\phi_{j2,\text{edge}}\coloneqq N_{\text{edge}}^{-1/2}(-1)^j\sinh(\lambda (L-j+1)),\\
&N_{\text{edge}}\coloneqq \frac{\sinh \lambda(2L+1)}{2\sinh \lambda}-L-\frac{1}{2},\\
&\epsilon_{\text{edge}}\coloneqq\frac{\sinh(\lambda)}{\sinh(L\lambda)}, 
\end{align}
where $\lambda$ is the solution of
\begin{align}
\frac{\sinh(\lambda(L+1))}{\sinh(\lambda L)}=\frac{t}{\mu}\label{eqlambda}
\end{align}
and is approximately given by $\lambda = \log(t/\mu)$ for large $L$~[Fig.~\ref{fig3}(b)]. The energy eigenvalue of the edge mode is exponentially small,
\begin{align}
&\epsilon_{\text{edge}}=\frac{\sinh(\lambda)}{\sinh(L\lambda)}\simeq\frac{t^2-\mu^2}{t}\Big(\frac{\mu}{t}\Big)^{L}, 
\end{align}

The ground state occupies all negative energy states:
\begin{align}
|\Phi_0\rangle=\hat{\tilde{\gamma}}_{\text{edge}}^\dagger\prod_{n=1}^L\hat{\tilde{\gamma}}_{n}^\dagger|0\rangle.
\end{align}
In addition, there are three quasi-degenerate ground states:
\begin{align}
&|\Phi_1\rangle=\hat{\gamma}_{\text{edge}}^\dagger|\Phi_0\rangle,\\
&|\Phi_2\rangle=\hat{\tilde{\gamma}}_{\text{edge}}|\Phi_0\rangle,\\
&|\Phi_3\rangle=\hat{\tilde{\gamma}}_{\text{edge}}\hat{\gamma}_{\text{edge}}^\dagger|\Phi_0\rangle.
\end{align}
These ground states are degenerate in the large $L$ limit, but possess an exponentially small energy splitting of the size $\epsilon_{\text{edge}}$, $\epsilon_{\text{edge}}$, $2\epsilon_{\text{edge}}$ respectively.

\subsection{Frustration-free decomposition}
Regardless of the presence of these edge modes, the Hamiltonian can also be written in a frustration-free form:
\begin{align}
\hat{H}&=\sum_{j=1}^L\hat{H}_j+E_0,\\
\hat{H}_j&=\sum_{\sigma=1}^2\big(\hat{\psi}_{j\sigma}^\dagger\hat{\psi}_{j\sigma}+\hat{\tilde{\psi}}_{j\sigma}\hat{\tilde{\psi}}_{j\sigma}^\dagger\big).
\end{align}
Here, $\hat{\psi}_{j\sigma}$ and $\hat{\tilde{\psi}}_{j\sigma}$ are defined by
\begin{align}
&\hat{\psi}_{j\sigma}\coloneqq\sum_{j'=1}^L\sum_{\sigma'=1}^2\Big(\sqrt{h^{(+)}}\Big)_{j\sigma,j'\sigma'}\hat{c}_{j'\sigma'},\\
&\hat{\tilde{\psi}}_{j\sigma}\coloneqq\hat{\Gamma}\hat{\psi}_{j\sigma}\hat{\Gamma}^\dagger,
\end{align}
where
\begin{align}
\Big(\sqrt{h^{(+)}}\Big)_{j\sigma,j'\sigma'}&\coloneqq\sum_{n=1}^{L-1}\sqrt{\epsilon_{n}}\phi_{j\sigma,n}\phi_{j'\sigma',n}\notag\\
&\quad+\sqrt{\epsilon_{\text{edge}}}\phi_{j\sigma,\text{edge}}\phi_{j'\sigma',\text{edge}}.
\end{align}
Since all bulk modes are gapped and the contribution to $\sqrt{h^{(+)}}$ from the edge modes is exponentially suppressed, the matrix elements $\Bigl(\sqrt{h^{(+)}}\Bigr)_{j\sigma,j'\sigma'}$ exhibit exponential decay of the form
\begin{align}
\Big|\Big(\sqrt{h^{(+)}}\Big)_{j\sigma,j'\sigma'}\Big|\leq Ce^{-A|i-j|},\quad A\coloneqq\log(\mu/t),
\end{align}
which we confirm numerically in Fig.~\ref{fig3}(c,d). 

\section{Relation to alternative approach}
\label{Kitaev}
Let us now discuss the relation of our results to the more general arguments of Ref.~\onlinecite{KITAEV20062}, which are applicable even to interacting systems.
Let $|\Phi_0\rangle,|\Phi_1\rangle,|\Phi_2\rangle,\dots$ be the eigenstates of the Hamiltonian $\hat{H}\coloneqq\sum_{i}\hat{H}_{i}$, labeled by the corresponding eigenvalues $E_0\leq E_1\leq E_2\leq\dots$. We assume the energy gap $E_1 - E_0 \geq \Delta$ remains nonzero in the large-$L$ limit. Without loss of generality, we set $E_0=0$.

In Appendix~D.1.2 of Ref.~\onlinecite{KITAEV20062}, an algorithm is given that modifies each $\hat{H}_i$ to a new operator $\hat{\tilde{H}}_{i}$ so that the ground state $|\Phi_0\rangle$ is  a zero-energy eigenstate of every $\hat{\tilde{H}}_{i}$ simultaneously, while leaving the total Hamiltonian unchanged, i.e., $\hat{H}=\sum_{i}\hat{\tilde{H}}_{i}$. 
Specifically, such an operator $\hat{\tilde{H}}_{i}$ is obtained by the integral
\begin{align}
\hat{\tilde{H}}_{i}&\coloneqq \int_{-\infty}^\infty dt w(t)e^{i\hat{H}t}\hat{H}_{i}e^{-i\hat{H}t}.
\end{align}
Here, $w(t)$ is a real function whose Fourier transform $\tilde{w}(\epsilon)\coloneqq\int_{-\infty}^\infty dt\,w(t)\,e^{i\epsilon t}$ $\tilde{w}(\epsilon)\coloneqq\int_{-\infty}^\infty dt w(t)e^{i\epsilon t}$ is (i) normalized ($\tilde{w}(\epsilon=0)=1$), (ii) compactly supported (i.e., $\tilde{w}(\epsilon)=0$ for $|\epsilon|>\Delta$), and (iii) smooth (i.e., infinitely differentiable).
By definition, the matrix elements of $\hat{\tilde{H}}_{i}$ read 
\begin{align}
\langle\Phi_n|\hat{\tilde{H}}_{i}|\Phi_m\rangle=\tilde{w}(E_n-E_m)\langle\Phi_n|\hat{H}_{i}|\Phi_m\rangle\label{elements}
\end{align}
The filtered operator $\hat{\tilde{H}}_i$ is no longer finite-ranged but the operator norm $\|[\hat{\tilde{H}}_i,\hat{\sigma}_j^a]\|$ decays faster than any power of $|i-j|$ (Appendix~\ref{a1}).
Recently, this argument was extended to systems that exhibit quasi-degeneracy with an Ising symmetry, where $E_1 - E_0$ decays exponentially with system size, but $E_2 - E_1$ remains finite and thus provides the relevant gap $\Delta$~\cite{Sahay2025EnforcedGaplessness}.

\begin{figure}[t]
\begin{center}
\includegraphics[width=\columnwidth]{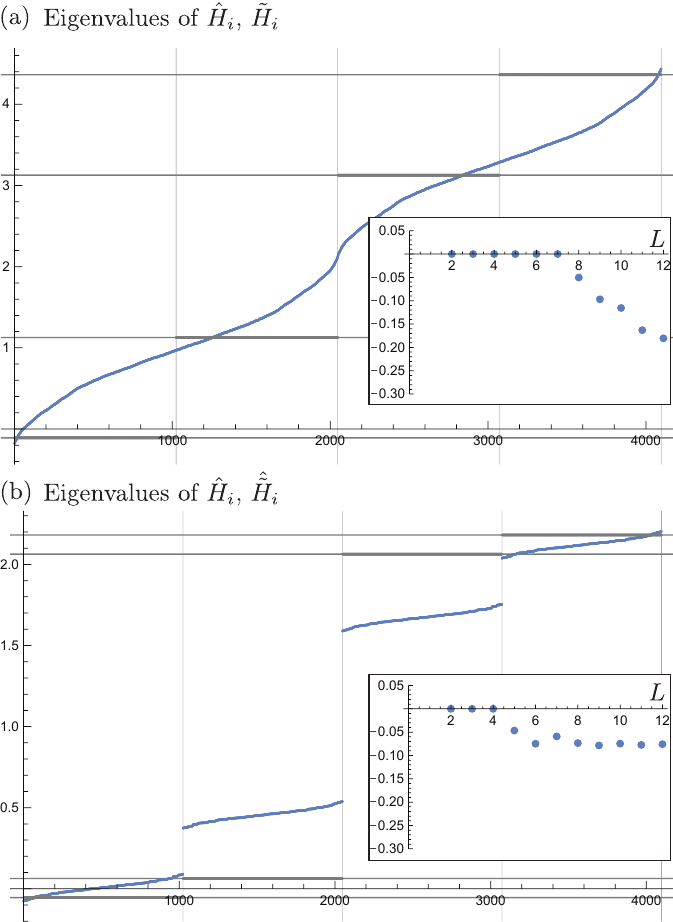}
\end{center}
\caption{
Eigenvalues of $\hat{H}_i$ (gray) and $\hat{\tilde{H}}_i$ (blue) for $L=12$ obtained by exact diagonalization.
Panel (a) shows results for $B=2$, while panel (b) corresponds to $B=0.5$. 
The inset plots the minimum eigenvalue of $\hat{\tilde{H}}_i$ as a function of $L$.
\label{fig4}
}
\end{figure}

Although such a decomposition resembles that for a frustration-free Hamiltonian, in general each $\hat{\tilde{H}}_{i}$ may fail to be positive-semidefinite, and $|\Phi_0\rangle$ might not be its ground state. As an example, let us consider the transverse-field Ising model under periodic boundary conditions:
\begin{align}
&\hat{H}=\sum_{i=1}^L\hat{H}_i,\\
&\hat{H}_i=-\hat{\sigma}_i^x\sigma_{i+1}^{x}-\frac{B}{2}(\hat{\sigma}_i^z+\hat{\sigma}_{i+1}^z)-e_0\quad (B>0).\label{Ising}
\end{align}
The ground state energy per spin~\cite{PhysRevB.35.7062,10.21468/SciPostPhysLectNotes.82}
\begin{align}
e_0\coloneqq-\frac{1}{L}\sum_{m=1}^L\sqrt{1+B^2+2B\cos\Big(\frac{2\pi}{L}\Big(m-\frac{1+(-1)^L}{4}\Big)\Big)}
\end{align}
is subtracted from $\hat{H}_i$ so that $E_0=0$. In the large-$L$ limit, one finds $e_0=-\frac{2(1+B)}{\pi}E(\frac{4B}{(1+B)^2})$, where $E(x)\coloneqq\int_0^{\frac{\pi}{2}}d\theta\sqrt{1-x\sin^2\theta}$ is the complete elliptic integral. 
The eigenvalues of $\hat{H}_i$ are $\pm\sqrt{1+B^2}-e_0$ and $\pm1-e_0$.

\begin{figure}[t]
\begin{center}
\includegraphics[width=\columnwidth]{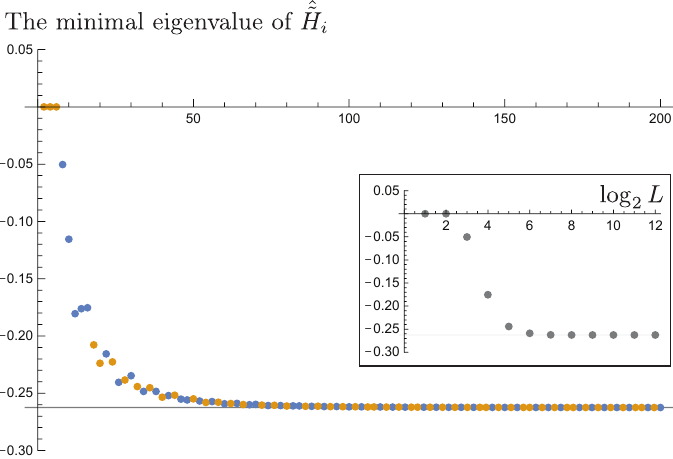}
\end{center}
\caption{
The minimum eigenvalue of $\hat{\tilde{H}}_i$ at $B=2$ obtained by mapping to free fermions by the Jordan-Wigner transformation. Orange (blue) points correspond to even (odd) fermion parity. The inset is for larger values of $L$.
\label{fig5}
}
\end{figure}

We set $\Delta=2(B-1)$ in the disordered phase ($B>1$) and $\Delta=4(1-B)$ in the ordered phase ($0<B<1$). Moreover, we define $\tilde{w}(\epsilon)$ to be the following bump function:
\begin{align}
\tilde{w}(\epsilon)=\begin{cases}
\exp\big(1+\frac{\Delta^2}{\epsilon^2-\Delta^2}\big)&(|\epsilon|<\Delta)\\
0&(|\epsilon|\geq\Delta)
\end{cases},\label{bump}
\end{align}
whose corresponding $w(t)$ is discussed in the Appendix~\ref{a1}. We numerically calculate the eigenvalues of $\hat{\tilde{H}}_{i}$ via Eq.~\eqref{elements} by exact diagonalization and present the results in Fig.~\ref{fig4}. Clearly, $\hat{\tilde{H}}_{i}$ can have negative eigenvalues and thus fails to be positive-semidefinite. To confirm that the minimum eigenvalue of $\hat{\tilde{H}}_{i}$ remains negative for larger values of $L$, we map the model to a free fermion by the Jordan-Wigner transformation (see Appendix~\ref{a2}). We show our results from this approach in Fig.~\ref{fig5}, which clearly indicate the convergence of the the minimum eigenvalue to $-0.262498$ in the large $L$ limit. In the Appendix~\ref{a2}, we also performed the same analysis for different choices of $\hat{H}_i$.

\section{Conclusion}

Appendix~D of Kitaev’s celebrated work~\cite{KITAEV20062} contains two profound observations concerning frustration-free systems. 
Applying the general arguments on free Majorana fermions to tight-binding models of complex fermions, we have constructed three explicit frustration-free, free-fermion models of the form $\hat{H}=\sum_i\hat{H}_i$, each displaying behaviors previously deemed unachievable in this class. These include a system with a nonzero Chern number, another with quasi-degenerate ground states whose finite-size splitting persists, and a third with gapless excitations that exhibit finite-size gaps scaling as $1/L$.

In all three cases, the individual terms $\hat{H}_i$ possess long tails---exponential in the first two and algebraic in the third. Crucially, however, this modification affects only the decomposition of the Hamiltonian: the total Hamiltonian $\hat{H}=\sum_i\hat{H}_i$ itself remains composed entirely of finite-range interactions and hoppings.

We also revisited the second, more general argument in the same appendix~\cite{KITAEV20062}. There, we pointed out that the modified local terms 
$\hat{\tilde{H}}_i$ need not be positive-semidefinite. 
We note that, in the limit of vanishing filtering gap $\Delta \rightarrow 0$, positive-semidefiniteness of the transformed local Hamiltonian terms may be restored. However, as the finiteness of the gap $\Delta$ controls the spatial decay of the transformed interactions, this observation does not establish the existence of a frustration-free decomposition with exponentially decaying interactions.
Consequently, this argument does not rigorously establish frustration-freeness in the broader setting considered there.

Together, these findings offer new insights into both the potential and limitations of frustration-free systems, helping to refine our understanding of their general structure and physical realizability. As a natural direction for future research, it is important to rigorously establish or disprove the refined conjecture concerning the incompatibility between chiral phases and gapped frustration-free Hamiltonians composed of strictly finite-ranged local terms. Heuristically, if such a model under periodic boundary conditions were to exist, one could remove terms across the boundary to reinterpret the system as an open-boundary model while preserving frustration-freeness. This would conflict with the gap scaling constraints established under open boundary conditions, such as those in Ref.~\onlinecite{10.1063/1.5089773}. While this heuristic supports the conjecture, a fully rigorous justification remains an important open problem.

\begin{acknowledgments}
We thank Ken Shiozaki, Yohei Fuji, Rintaro Masaoka, Seishiro Ono, Gil Young Cho, Carolyn Zhang, and Ruben Verresen for useful discussions.
The work of H.W. is supported by JSPS KAKENHI Grant No.~JP24K00541.
H.C.P.~acknowledges support from the National Key R\&D Program of China (Grant No. 2021YFA1401500) and the Hong Kong Research Grants Council (C7037-22GF).
\end{acknowledgments}

\appendix
\section{Bump function and its Fourier transform}
\label{a1}
Here we summarize the properties of $w(t)$ corresponding to the Fourier transform $\tilde{w}(\epsilon)$ in Eq.~\eqref{bump}. Note, however, that our discussion in Sec.~\ref{Kitaev} does not rely on the real-time formula. The asymptotic behavior of $w(t)$ can be derived via the saddle-point approximation~\cite{johnson2015saddle}:
\begin{align}
|w(t)|\simeq \Delta \sqrt{\pi} \Big(\frac{2e}{t\Delta}\Big)^{3/4}  e^{-\sqrt{t\Delta}}\cos\Big(t\Delta-\sqrt{t\Delta}-\frac{3\pi}{8}\Big).
\end{align}
We thus find
\begin{align}
|w(t)|\leq 8\Delta\frac{e^{-\sqrt{t\Delta}}}{(t\Delta)^{3/4}}
\end{align}
for any $t$, which decays faster than any power of $1/(t\Delta)$. Indeed, for any $t \in \mathbb{R}$ and $n=0,1,2,3,\dots$, we have
\begin{align}
|w(t)| \,\le\, c_n\,\Delta\,(t\Delta)^{-n},\label{bound}
\end{align}
where, for instance, one may take $c_n = 8\,n^{2n}$ for $n=1,2,3,\dots$ and $c_0=8$. Other examples of $\omega(t)$ exhibiting even faster decay can be found in Ref.~\onlinecite{Dziubanski_Hernandez_1998}.

Using Eq.~\eqref{bound}, one can discuss the locality of the filtered local operator $\hat{\tilde{H}}_{i}$~\cite{hastings2010locality}. To this end, let us compute
\begin{align}
\|[\hat{\tilde{H}}_{i},\hat{O}_{j}]\|\leq \int_{-\infty}^\infty dt w(t)\big\|[e^{i\hat{H}t}\hat{H}_{i}e^{-i\hat{H}t},\hat{O}_{j}]\big\|
\end{align}
for an operator $\hat{O}_{j}$ supported around $j$.  The Lieb-Robinson bound states that
\begin{align}
\big\|[e^{i\hat{H}t}\hat{H}_{i}e^{-i\hat{H}t},\hat{O}_{j}]\big\|\leq 2\|\hat{H}_{i}\|\|\hat{O}_{j}\|\min\Big\{e^{-\frac{|i-j|-v|t|}{\xi_0}},1\Big\}
\end{align}
for a constant $\xi_0$ and the Lieb-Robinson velocity $v$. It follows that
\begin{align}
&\|[\hat{\tilde{H}}_{i},\hat{O}_{j}]\|\leq  2\|\hat{H}_{i}\|\|\hat{O}_{j}\|\int_{|t|<\frac{|i-j|}{2v}} dt w(t)e^{-\frac{|i-j|-v|t|}{\xi_0}}\notag\\
&\quad\quad\quad\quad\quad\quad+2\|\hat{H}_{i}\|\|\hat{O}_{j}\|\int_{|t|>\frac{|i-j|}{2v}} dt w(t)\notag\\
&\leq  2\|\hat{H}_{i}\|\|\hat{O}_{j}\|\int_{|t|<\frac{|i-j|}{2v}} dt c_0\Delta e^{-\frac{|i-j|-v|t|}{\xi_0}}\notag\\
&\quad+2\|\hat{H}_{i}\|\|\hat{O}_{j}\|\int_{|t|>\frac{|i-j|}{2v}} dt c_{n+1}\,\Delta\,(t\Delta)^{-(n+1)}\notag\\
&\leq  4\|\hat{H}_{i}\|\|\hat{O}_{j}\|\Big(c_0\frac{\xi_0\Delta}{v}e^{-\frac{|i-j|}{2\xi_0}}+\frac{c_{n+1}}{n}\Big(\frac{2v}{|i-j|\Delta}\Big)^{n}\Big),
\end{align}
implying that $\|[\hat{\tilde{H}}_{i},\hat{O}_{j}]\|$ decays faster than any power of $\frac{v}{|i-j|\Delta}$.

\section{Negative eigenvalues of $\hat{\tilde{H}}_j$ in the transverse--field Ising chain}
\label{a2}
\subsection{The Hamiltonian and its spectrum}
The Hamiltonian of the transverse--field Ising chain under periodic boundary conditions is
\begin{align}
  \hat H &\coloneqq
  -\sum_{j=1}^{L}\hat{\sigma}_j^{x}\hat{\sigma}_{j+1}^{x}
  -B\sum_{j=1}^{L}\hat{\sigma}_j^{z}
  -L e_0.
\end{align}
The constant $e_0$ is fixed so that the shifted ground--state energy is $E_0=0$.
To simplify the discussion we focus on the disordered phase ($B>1$).
Also, for a technical reason we assume $L$ is even.

Introduce spinless fermions by
\begin{align}
  \hat{c}_j &\coloneqq\Big(\prod_{l=1}^{j-1}\hat{\sigma}_l^z\Big)\frac{\hat{\sigma}_j^{x}+i\hat{\sigma}_j^{y}}{2},\\
  \hat{c}_j^\dagger &\coloneqq\Big(\prod_{l=1}^{j-1}\hat{\sigma}_l^z\Big)\frac{\hat{\sigma}_j^{x}-i\hat{\sigma}_j^{y}}{2}.
\end{align}
Then the Pauli operators can be written as
\begin{align}
\begin{aligned}
\hat{\sigma}_j^z &= 1 - 2\hat{c}_j^\dagger \hat{c}_j,\\
\hat{\sigma}_j^x &= \prod_{l=1}^{j-1}( 1 - 2\hat{c}_l^\dagger \hat{c}_l)(\hat{c}_j + \hat{c}_j^\dagger),
\end{aligned}
\end{align}
We find
\begin{align}
\hat{\sigma}_j^x\hat{\sigma}_{j+1}^x&=(\hat{c}_j + \hat{c}_j^\dagger)( 1 - 2\hat{c}_j^\dagger \hat{c}_j)(\hat{c}_{j+1} + \hat{c}_{j+1}^\dagger)\notag\\
&=(-\hat{c}_j+\hat{c}_j^\dagger)(\hat{c}_{j+1} + \hat{c}_{j+1}^\dagger)\notag\\
&=\hat{c}_j^\dagger \hat{c}_{j+1}+\hat{c}_{j+1}^\dagger \hat{c}_j + \hat{c}_j^\dagger \hat{c}_{j+1}^\dagger+\hat{c}_{j+1}\hat{c}_j.
\end{align}

\begin{align}
\hat{\sigma}_L^x\hat{\sigma}_1^x&=(-1)^{\sum_{l=1}^{L}\hat{c}_l^\dagger \hat{c}_l}(1 - 2\hat{c}_L^\dagger \hat{c}_L)(\hat{c}_L + \hat{c}_L^\dagger)(\hat{c}_1 + \hat{c}_1^\dagger)\notag\\
&=-\hat{P}(-\hat{c}_L+\hat{c}_L^\dagger )(\hat{c}_1+ \hat{c}_1^\dagger)\notag\\
&=-\hat{P}(\hat{c}_L^\dagger \hat{c}_1 +\hat{c}_1^\dagger \hat{c}_L + \hat{c}_L^\dagger \hat{c}_1^\dagger+\hat{c}_1\hat{c}_L ),
\end{align}
where $\hat{P}=(-1)^{\sum_{j=1}^{L}\hat{c}_j^\dagger \hat{c}_j}$ is the fermion parity.
Hence, the total Hamiltonian is
\begin{align}
\hat H&=-\sum_{j=1}^{L-1}(\hat{c}_j^\dagger \hat{c}_{j+1}+\hat{c}_{j+1}^\dagger \hat{c}_j + \hat{c}_j^\dagger \hat{c}_{j+1}^\dagger+\hat{c}_{j+1}\hat{c}_j)\notag\\
&\quad+\hat{P}(\hat{c}_L^\dagger \hat{c}_1 +\hat{c}_1^\dagger \hat{c}_L + \hat{c}_L^\dagger \hat{c}_1^\dagger+\hat{c}_1\hat{c}_L)\notag\\
&\quad+\sum_{j=1}^{L}B(\hat{c}_j^\dagger \hat{c}_j-\hat{c}_j \hat{c}_j^\dagger)-Le_0.
\end{align}

We define the Fourier transformation by
\begin{align}
\hat{c}_k^\dagger \coloneqq\frac{1}{\sqrt{L}}\sum_{j=1}^Le^{ikj}\hat{c}_j^\dagger.
\end{align}
for $k\in K_P$, where
\begin{align}
K_P=
\begin{cases}
\big\{\frac{2\pi}{L}m\big\}_{m=1-L/2}^{L/2}&\text{in the $\hat{P}=-1$ sector},\\
\big\{\frac{2\pi}{L}(m-\frac{1}{2})\big\}_{m=1-L/2}^{L/2}&\text{in the $\hat{P}=+1$ sector}.\\
\end{cases}
\end{align}
We find
\begin{align}
\hat{H}&=\sum_{k\in K_P}\hat{H}_k-Le_0,\\
\hat{H}_k&\coloneqq(B-\cos k)\hat{c}_k^\dagger \hat{c}_{k}-(B-\cos k) \hat{c}_{-k} \hat{c}_{-k}^\dagger\notag\\
&\quad- i\sin k (\hat{c}_k^\dagger \hat{c}_{-k}^\dagger -\hat{c}_{-k}\hat{c}_k),
\end{align}

The odd parity sector $K_{P=-1}$ contains both $k=0$ and $k=\pi$.  We treat these points separately as they are already diagonal.
\begin{align}
\hat{H}_0&=(B-1)(2\hat{c}_0^\dagger \hat{c}_0-1),\\
\hat{H}_\pi&=(B+1)(2\hat{c}_\pi^\dagger \hat{c}_{\pi}-1)
\end{align}

For $k\in K_P$ not equal to $0$ or $\pi$, we group $\hat{H}_k$ and $\hat{H}_{-k}=\hat{H}_{k}$ and write
\begin{align}
&\hat{H}_k+\hat{H}_{-k}\notag\\
&=2\begin{pmatrix} \hat{c}_k^\dagger & \hat{c}_{-k} \end{pmatrix}
\begin{pmatrix}
B - \cos k & -i\sin k \\
i\sin k & -(B - \cos k)
\end{pmatrix}
\begin{pmatrix} \hat{c}_k \\ \hat{c}_{-k}^\dagger \end{pmatrix}\notag\\
&=2\begin{pmatrix} \hat{\gamma}_k^\dagger & \hat{\gamma}_{-k} \end{pmatrix}
\begin{pmatrix}
\epsilon_k &0 \\
0 & -\epsilon_k
\end{pmatrix}
\begin{pmatrix} \hat{\gamma}_k \\ \hat{\gamma}_{-k}^\dagger \end{pmatrix}\notag\\
&=2\epsilon_k (\hat{\gamma}_k^\dagger \hat{\gamma}_k + \hat{\gamma}_{-k}^\dagger \hat{\gamma}_{-k}-1),
\end{align}
where
\begin{align}
\begin{pmatrix} \hat{\gamma}_k \\ \hat{\gamma}_{-k}^\dagger \end{pmatrix}
\coloneqq
\begin{pmatrix}
\cos(\tfrac{\theta_k}{2}) & -i\sin(\tfrac{\theta_k}{2}) \\
-i\sin(\tfrac{\theta_k}{2}) & \cos(\tfrac{\theta_k}{2})
\end{pmatrix}
\begin{pmatrix} \hat{c}_k \\ \hat{c}_{-k}^\dagger \end{pmatrix},
\end{align}
\begin{align}
\cos(\tfrac{\theta_k}{2}) &\coloneqq \sqrt{\frac{1}{2} \left(1 + \frac{B - \cos k}{\epsilon_k}\right)}, \\
\sin(\tfrac{\theta_k}{2}) &\coloneqq\text{sign}(\sin k)\sqrt{\frac{1}{2} \left(1 - \frac{B - \cos k}{\epsilon_k}\right)},
\end{align}
and 
\begin{align}
\epsilon_k \coloneqq \sqrt{1+B^2 - 2B\cos k}.
\end{align}
Since $\theta_{-k}=-\theta_k$ and $\epsilon_{-k}=\epsilon_k$, the definition $\hat{\gamma}_k =\cos(\tfrac{\theta_k}{2})\hat{c}_k - i\sin(\tfrac{\theta_k}{2})\hat{c}_{-k}^\dagger$ is applicable when $k$ is replaced by $-k$. 
We have
\begin{align}
\sin\theta_k
= \frac{\sin k}{\epsilon_k},
\quad
\cos\theta_k
= \frac{B - \cos k}{\epsilon_k}.
\end{align}
As we assume $B>1$, $\epsilon_{0}=B-1,\theta_0=0$ and  $\epsilon_{\pi}=B+1, \theta_{\pi}=0$.
After all, one can combine all terms and 
\begin{align}
\hat{H}&=\sum_{k\in K_P}2\epsilon_k\hat{\gamma}_k^\dagger \hat{\gamma}_k-\sum_{k\in K_P}\epsilon_k-Le_0.
\end{align}

The ground state in the positive parity sector is
\begin{align}
|\Phi_0\rangle\coloneqq N_0\prod_{k\in K_{P=+1}}\hat{\gamma}_k|0\rangle,
\end{align}
where $N_0>0$ is the normalization factor.
The ground state energy  is
\begin{align}
E_0=-Le_0-\sum_{k\in K_{P=+1}}\epsilon_k=-Le_0-\sum_{m=1-L/2}^{L/2}\epsilon_{k=\frac{2\pi}{L}(m-\frac{1}{2})}
\end{align}
so that we identify
\begin{align}
e_0=-\frac{1}{L}\sum_{m=1-L/2}^{L/2}\epsilon_{k=\frac{2\pi}{L}(m-\frac{1}{2})}
\end{align}
assuming that the ground state in the other sector has a higher energy, which is indeed the case in this model.

On the other hand, the ground state in the odd parity sector (which contains both $k=0,\pi$) is
\begin{align}
|\Phi_1\rangle\coloneqq\hat{c}_0^\dagger|\tilde{\Phi}_0\rangle,
\end{align}
where
\begin{align}
|\tilde{\Phi}_0\rangle\coloneqq\tilde{N}_0\prod_{k\in K_{P=-1},k\neq0,\pi}\hat{\gamma}_k|0\rangle
\end{align}
and $\tilde{N}_0>0$ is the normalization factor.  The ground state energy is
\begin{align}
E_1&=-Le_0-\sum_{k\in K_{P=-1}}\epsilon_k+2\epsilon_0=2(B-1)+\delta,\\
\delta&\coloneqq\Big(\sum_{m=1-L/2}^{L/2}\epsilon_{k=\frac{2\pi}{L}(m-\frac{1}{2})}\Big)-\Big(\sum_{m=1-L/2}^{L/2}\epsilon_{k=\frac{2\pi}{L}m}\Big).
\end{align}
Note that the quantity $\delta$ is exponentially small with the system size.

Single-particle excited states, labeled by $k\in K_{P=-1}$ including $k=0,\pi$, and their energy eigenvalues can be written as
\begin{align}
|k\rangle&\coloneqq\hat{\gamma}_k^\dagger|\tilde{\Phi}_0\rangle,\\
E_k&\coloneqq E_1+2(\epsilon_k-\epsilon_0)=2\epsilon_k+\delta.
\end{align}
In particular, we identify the gap
\begin{align}
E_1-E_0>\Delta\coloneqq2(B-1).
\end{align}

\subsection{Local Hamiltonians}
Now we expand the local Hamiltonian using quasi-particle operators.
\begin{align}
\hat{c}_j &=\frac{1}{\sqrt{L}}\sum_{q'\in K_{P}}e^{iq'j}\Big(\cos(\tfrac{\theta_{q'}}{2})\hat{\gamma}_{q'}  +i\sin(\tfrac{\theta_{q'}}{2} )\hat{\gamma}_{-{q'}}^\dagger\Big),\\
\hat{c}_j^\dagger &=\frac{1}{\sqrt{L}}\sum_{q\in K_{P}}e^{-iqj}\Big(\cos(\tfrac{\theta_q}{2})\hat{\gamma}_q^\dagger  -i\sin(\tfrac{\theta_q}{2} )\hat{\gamma}_{-q}\Big).
\end{align}
We keep only terms that keeps the number of quasi-particles (i.e., terms of the form $\hat{\gamma}_k^\dagger\hat{\gamma}_{k'}$ and $\hat{\gamma}_{k'}\hat{\gamma}_k^\dagger$):
\begin{align}
\hat{c}_j^\dagger \hat{c}_l&=\sum_{k,k'\in K_{P}}\frac{1}{L}
e^{-ikj+ik'l}\cos(\tfrac{\theta_k}{2})\cos(\tfrac{\theta_{k'}}{2})\hat{\gamma}_k^\dagger\hat{\gamma}_{k'}\notag\\
&\quad+\sum_{k,k'\in K_{P}}\frac{1}{L}e^{-ikl+ik'j}\sin(\tfrac{\theta_{k'}}{2})\sin(\tfrac{\theta_{k}}{2})\hat{\gamma}_{k'}\hat{\gamma}_{k}^\dagger\notag\\
&\quad+\text{($\hat{\gamma}_{-k}\hat{\gamma}_{k'}$ and $\hat{\gamma}_k^\dagger\hat{\gamma}_{-k'}^\dagger$ terms)}.
\end{align}
\begin{align}
\hat{c}_j^\dagger \hat{c}_l^\dagger&=\sum_{k,k'\in K_{P}}\frac{1}{L}
e^{-ikj+ik'l}\hat{\gamma}_k^\dagger\hat{\gamma}_{k'}i\cos(\tfrac{\theta_k}{2})\sin(\tfrac{\theta_{k'}}{2})\notag\\
&\quad+\sum_{k,k'\in K_{P}}\frac{1}{L}e^{-ikl+ik'j}\hat{\gamma}_{k'}\hat{\gamma}_{k}^\dagger i\cos(\tfrac{\theta_k}{2})\sin(\tfrac{\theta_{k'}}{2})\notag\\
&\quad+\text{($\hat{\gamma}_{-k}\hat{\gamma}_{k'}$ and $\hat{\gamma}_k^\dagger\hat{\gamma}_{-k'}^\dagger$ terms)}.
\end{align}

\begin{table*}
\begin{center}
\caption{The minimum eigenvalue of $\hat{\tilde{H}}_j^{(l)}$ at $B=2$. \label{tab}}
\begin{tabular}{c|ccccc} \hline
 & $L=8$ & $L=10$ &$L=12$ & $L=1024$ & $L=4096$ \\ \hline
$\hat{\tilde{H}}_j^{(1)}$&
$-0.0503557287813$ & 
$-0.1155200624828$ & 
$-0.1805373136059$ & 
$-0.262497894167$ & 
$-0.262497894223$ \\ 
(ED) & 
$-0.0503557287813$ & 
$-0.1155200624829$ & 
$-0.1805373136062$ & NA & NA \\ \hline
$\hat{\tilde{H}}_j^{(2)}$& 
$-0.0590798054631$ & 
$-0.1443947637723$ & 
$-0.1910645333616$ & 
$-0.281507712689$ & 
$-0.281507712753$ \\ 
(ED) & 
$-0.0590798054631$ & 
$-0.1443947637725$ & 
$-0.1910645333616$ & NA & NA \\ \hline
$\hat{\tilde{H}}_j^{(3)}$& 
$-0.0599187038609$ & 
$-0.1500592435439$ & 
$-0.1928907968755$ & 
$-0.283754747078$ & 
$-0.283754747143$ \\ 
(ED) & 
$-0.0599187038609$ & 
$-0.1500592435440$ & 
$-0.1928907968763$& NA & NA \\ \hline
 \end{tabular}
 \end{center}
\end{table*}

For the choice of the local Hamiltonian used in the main text
\begin{align}
\hat H_j^{(1)}&\coloneqq-\hat{\sigma}_j^x\hat{\sigma}_{j+1}^x  -\frac{B}{2}(\hat{\sigma}_j^{z}+\hat{\sigma}_{j+1}^{z})  - e_0\notag\\
&=-\hat{c}_j^\dagger \hat{c}_{j+1}-\hat{c}_{j+1}^\dagger \hat{c}_j - \hat{c}_j^\dagger \hat{c}_{j+1}^\dagger-\hat{c}_{j+1}\hat{c}_j\notag\\
&\quad+B(\hat{c}_j^\dagger \hat{c}_j+\hat{c}_{j+1}^\dagger \hat{c}_{j+1})
-(B+e_0),
\end{align}
we find
\begin{align}
\hat{H}_j^{(1)}&=\sum_{k,k'\in K_{P}}e^{i(k'-k)j}h_{k,k'}^{(1)}\hat{\gamma}_k^\dagger\hat{\gamma}_{k'}+\frac{\delta}{L}\delta_{P,-1}\notag\\
&\quad+\text{($\hat{\gamma}_{-k}\hat{\gamma}_{k'}$ and $\hat{\gamma}_k^\dagger\hat{\gamma}_{-k'}^\dagger$ terms)},\\
h_{k,k'}^{(1)}&\coloneqq-\frac{2}{L}e^{i(k'-k)/2}\cos(\tfrac{k+k'}{2}+\tfrac{\theta_k+\theta_{k'}}{2})\notag\\
&\quad+\frac{2}{L}e^{i(k'-k)/2}B\cos(\tfrac{k-k'}{2})\cos(\tfrac{\theta_k+\theta_{k'}}{2}).
\end{align}

For a different choice of the local term
\begin{align}
\hat H_j^{(2)}&\coloneqq-\frac{1}{2}(\hat{\sigma}_{j-1}^{x}\hat{\sigma}_{j}^{x}+\hat{\sigma}_{j}^{x}\hat{\sigma}_{j+1}^{x})-B\hat{\sigma}_{j}^{z}-e_{0}\notag\\
&=\frac{1}{2}(-\hat{c}_{j-1}^\dagger \hat{c}_j-\hat{c}_j^\dagger \hat{c}_{j-1} - \hat{c}_{j-1}^\dagger \hat{c}_j^\dagger-\hat{c}_j\hat{c}_{j-1} ) \notag\\
&\quad+\frac{1}{2}(-\hat{c}_j^\dagger \hat{c}_{j+1}-\hat{c}_{j+1}^\dagger \hat{c}_j - \hat{c}_j^\dagger \hat{c}_{j+1}^\dagger-\hat{c}_{j+1}\hat{c}_j )\notag\\
&\quad+2B\hat{c}_j^\dagger \hat{c}_j-(B+e_0),
\end{align}
we find
\begin{align}
\hat{H}_j^{(2)}&=\sum_{k,k'\in K_{P}}e^{i(k'-k)j}h_{k,k'}^{(2)}\hat{\gamma}_k^\dagger\hat{\gamma}_{k'}+\frac{\delta}{L}\delta_{P,-1}\notag\\
&\quad
+\text{($\hat{\gamma}_{-k}\hat{\gamma}_{k'}$ and $\hat{\gamma}_k^\dagger\hat{\gamma}_{-k'}^\dagger$ terms)},\\
h_{k,k'}^{(2)}&\coloneqq-\frac{2}{L}\cos(\tfrac{k-k'}{2})\cos(\tfrac{k+k'}{2}+\tfrac{\theta_k+\theta_{k'}}{2})\notag\\
&\quad+\frac{2}{L}B\cos(\tfrac{\theta_k+\theta_{k'}}{2}).
\end{align}

For yet another choice
\begin{align}
\hat H_j^{(3)}&\coloneqq-\hat{\sigma}_{j}^{x}\hat{\sigma}_{j+1}^{x}-B\hat{\sigma}_{j}^{z}-e_{0}\notag\\
&=-\hat{c}_j^\dagger \hat{c}_{j+1}-\hat{c}_{j+1}^\dagger \hat{c}_j - \hat{c}_j^\dagger \hat{c}_{j+1}^\dagger-\hat{c}_{j+1}\hat{c}_j \notag\\
&\quad+2B\hat{c}_j^\dagger \hat{c}_j-(B+e_0),
\end{align}
we find
\begin{align}
\hat{H}_j^{(3)}&=\sum_{k,k'\in K_{P}}e^{i(k'-k)j}h_{k,k'}^{(3)}\hat{\gamma}_k^\dagger\hat{\gamma}_{k'}+\frac{\delta}{L}\delta_{P,-1}\notag\\
&\quad+\text{($\hat{\gamma}_{-k}\hat{\gamma}_{k'}$ and $\hat{\gamma}_k^\dagger\hat{\gamma}_{-k'}^\dagger$ terms)},\\
h_{k,k'}^{(3)}&\coloneqq-\frac{2}{L}e^{i(k'-k)/2}\cos(\tfrac{k+k'}{2}+\tfrac{\theta_k+\theta_{k'}}{2})\notag\\
&\quad+\frac{2}{L}B \cos(\tfrac{\theta_k+\theta_{k'}}{2}).
\end{align}

\subsection{The spectrally filtered local operator}

Since contributions from terms of the form $\hat{\gamma}_{-k}\hat{\gamma}_{k'}$ and $\hat{\gamma}_k^\dagger\hat{\gamma}_{-k'}^\dagger$ are annihilated by the spectral filter, we arrive at the exact expression ($l=1,2,3$):
\begin{align}
\hat{\tilde{H}}_j^{(l)}&=\sum_{k,k'\in K_{P}}e^{i(k'-k)j}\tilde{h}_{k,k'}^{(l)}\hat{\gamma}_k^\dagger\hat{\gamma}_{k'}+\frac{\delta}{L}\delta_{P,-1},\\
\tilde{h}_{k,k'}^{(l)}&\coloneqq\tilde{w}(2(\epsilon_k-\epsilon_{k'}))h_{k,k'}^{(l)}.
\end{align}

We obtain the exact spectrum of $\hat{\tilde{H}}_j^{(l)}$ by diagonalizing $H_{k,k'}^{(l)}$. 
We summarize our results in Table~\ref{tab}.
We also confirm the validity of our calculations by exact diagonalization (ED) $L=8,10,12$.

\bibliography{refs}

\end{document}